%% file: main.tex
\documentclass[submission,copyright,creativecommons]{eptcs}
\providecommand{\event}{GandALF 2023}

\usepackage{iftex}

\ifpdf
  \usepackage{underscore}         % Only needed if you use pdflatex.
  \usepackage[T1]{fontenc}        % Recommended with pdflatex
\else
  \usepackage{breakurl}           % Not needed if you use pdflatex only.
\fi

\usepackage{preamble} 
\title{The Recursive Arrival Problem}
\author{Thomas Webster 
\institute{University of Edinburgh, UK}
\email{Thomas.Webster@ed.ac.uk}}

\def\titlerunning{The Recursive Arrival Problem}
\def\authorrunning{T. Webster}
\date{\today{}}

\begin{document}

\maketitle
\subfile{abs-intro}
\section{Preliminaries}\label{sec:prelim}
\subfile{arrival-introduction}
\subsection{The Recursive Arrival Problem}\label{sec:local-states}
\subfile{local-rec-arrival}

\newpage

\bibliographystyle{eptcs}
\bibliography{refs}
%\appendix
%\subfile{appendix}
\end{document}

%% file: abs-intro.tex
\begin{abstract}
We study an extension of the {\tt Arrival} problem, 
called {\tt Recursive Arrival}, inspired by Recursive State Machines, which allows for a family of switching graphs that can call each other in a recursive way. We study the computational complexity of deciding whether a Recursive Arrival instance terminates at 
a given target vertex.  We show this problem is contained in $\NP{}\cap\coNP{}$, and we show that a search version of the problem
lies in \UEOPL{}, and hence in $\EOPL = \PLS \cap \PPAD$. Furthermore, we show \P{}-hardness of the Recursive Arrival decision problem.  By contrast, the current best-known hardness result for {\tt Arrival} is $\PL$-hardness.
%\keywords{Arrival \and Recursive Markov Chains \and Recursive State Machines}
\end{abstract}

\section{Introduction}

{\tt Arrival} is a simply described decision problem defined by Dohrau, G\"{a}rtner, Kohler, Matou\u{s}ek and Welzl \cite{DGKMW16}.  Informally, it asks whether a train moving along the vertices of a given directed graph, with
$n$ vertices, will eventually reach a given target vertex, starting at
a given start vertex. At each vertex, $v$, there are (without loss of generality) two outgoing edges and the train moves deterministically, alternately taking the first out-edge, then the second, switching between them if and when it revisits that vertex repeatedly.   
This process is known as ``switching'' and can be viewed as a deterministic simulation of a random walk on the directed graph.  It can also be viewed
as a natural model of a state transition system where a local deterministic cyclic scheduler is
provided for repeated transitions out of each state. 

Dohrau et al. showed that this {\tt Arrival} decision problem
lies in the complexity class $\NP\cap\coNP$, but it is not known to be in \P{}. There has been much recent work showing that a search version of the {\tt Arrival} problem lies in subclasses of \TFNP{} including \PLS{} \cite{Kar17}, \CLS{} \cite{GHH+18}, and \UEOPL{} \cite{FGMS19}, as well as showing that {\tt Arrival} is in $\UP\cap\coUP$ \cite{GHH+18}. There has also been progress on lower bounds, including \PL{} hardness and \CC{} hardness \cite{Man21}. Further, another recent result by G\"{a}rtner et al. \cite{GHH21} gives an algorithm for {\tt Arrival} with running time $2^{\bigO(\sqrt{n}\log(n))}$, the first known sub-exponential algorithm. In addition, they give a polynomial-time algorithm for ``almost acyclic'' instances. Auger et al. also give a polynomial-time algorithm for instances on a ``tree-like multigraph'' \cite{ACD22}. 

The complexity of {\tt Arrival} is particularly interesting in the context of other games on graphs. For instance, Condon's simple stochastic games, mean-payoff games, and parity games \cite{Con92,ZP96,Jur98}, where the two-player variants are known to be in $\NP\cap\coNP$ and the one-player variants have polynomial time algorithms. {\tt Arrival}, however, is a zero-player game that has no known polynomial time algorithm and, furthermore, Fearnley et al. \cite{FGMS21} that a one-player generalisation of Arrival is, in fact, \NP{}-complete, in stark contrast to these two-player graph games.

We introduce and consider a new generalisation of Arrival that we call {\tt Recursive Arrival}, in which we are given a finite collection of Arrival instances (``components'') with the ability to, from certain nodes, invoke each other in a potentially recursive way. Each component has a set of entries and a set of exits, and we study the complexity of deciding whether the run starting from a given entry of a given component reaches a given exit of that component, which may involve recursive calls to other components. 

Our model is inspired by work on recursive transition systems started by Alur et al. \cite{AEY01} with Recursive State Machines (RSMs) modelling sequential imperative programming. These inspired further work on Recursive Markov Chains (RMCs), Recursive Markov Decision Processes (RMDPs), and Recursive Simple Stochastic Games (RSSGs) by Etessami and Yannakakis \cite{EY08, EY09, EY15}. RSMs
(and RMCs) are essentially ``equivalent'' (see 
\cite{EY09}) to (probabilistic) pushdown systems \cite{BEM97,EKM06}
and have applications in model-checking of procedural programs with recursion.
%Consideration of {\tt Recursive Arrival} is interesting within this context. 
%For example, the problem of deciding whether the value of Condon's %Simple Stochastic Games is at least one half is, like {\tt %Arrival}, known to be in $\NP{}\cap\coNP{}$ but not yet known to %be in \P{} \cite{Con92}.
%We know for RSSGs that restricting to just the single-exit positive-reward %case results in a harder problem not yet known to be in \PPAD{} and that %becomes undecidable with multiple exits.

There is previous work on Arrival generalisations including a variant we call {\tt Succinct Arrival}, where at a vertex $v$ the alternation takes the first outgoing edge of $v$ on the first $A_v$ visits and then the second edge the next $B_v$ visits, repeating this sequence indefinitely. The numbers $A_v$ and $B_v$ are given succinctly in binary as input, and hence may be exponentially larger than the bit encoding size of the instance. Fearnley et al. showed that {\tt Succinct Arrival} is \P{}-hard and in $\NP{}\cap\coNP{}$ \cite{FGMS21}.  However, we do not know any inter-reducibility between {\tt Recursive Arrival} and {\tt Succinct Arrival } variants.  In \cite{W22}, we also defined and studied a generalisation of {\tt Arrival} that allows both switching nodes as well as randomised nodes, and we showed that this results in \PP{}-hardness and containment in \PSPACE{} for (quantitative) reachability
problems.   

In this paper, we show that the {\tt Recursive Arrival} problem lies in $\NP{}\cap\coNP{}$, like {\tt Arrival}, by giving a generalised witness scheme that efficiently categorises both terminating and non-terminating instances. We also show that the  natural search version 
of {\tt Recursive Arrival} is in both \PLS{} and \PPAD{} and in fact in
\UEOPL{}, by giving a reduction to a canonical \UEOPL{} problem. 
We also show $\P{}$-hardness for the {\tt Recursive Arrival}
problem by reduction from the Circuit Value Problem.  This contrasts with the current best-known hardness
result for {\tt Arrival}, which is $\PL$-hardness (\cite{Man21}).

Due to space limitations, many proofs are relegated to the full version of the paper.

%% file: arrival-introduction.tex
Let $\nat = \{0,1,\ldots\}$ denote the natural numbers, and let $\natinf = \nat \cup \{ \infty \}$.  We assume the usual 
ordering on elements of $\natinf$. For $j \in \nat$ and $k \in \natinf$, we use the notion  $[j \ldots k] =\{i\in\nat \mid j\leq i\leq k\}$, and we define $[k] = [1 \ldots k]$. All propositions of this section follow directly from the cited prior works.

\begin{definition}\label{def:sim:arrival-graph}
A {\em switch graph} is given by a tuple $G:=(V,s^0,s^1)$ where, for each $\sigma\in\{0,1\}$, $s^\sigma:V\to V$ is a function from vertices to vertices. \qed
\end{definition}

Given a Switch Graph $G$, we define its directed {\em edges} to be the set $E:=\{(v,s^0(v)) \mid v\in V\}\cup\{(v,s^1(v))\mid v\in V\}$, with self-loops allowed. We write $E_\sigma:=\{(v,s^\sigma(v)) \mid v\in V\}$ for $\sigma\in\{0,1\}$ to refer to edges arising specifically from transitions $s^\sigma(v)$, for each vertex $v$.

Given a switch graph, $G:=(V,s^0,s^1)$, we say $q:V\to\{0,1\}$ is a {\em switch position} on $V$. We let $Q$ be the set of all switch positions on $V$ and define $q^0\in Q$ by $q^0(v)=0$ for all $v\in V$ as the {\em initial switch position}.
Given a switch graph, we say a {\em state} of the graph is an ordered pair $(v,q)\in V\times Q$ and we let $\Gamma = V \times Q$ be the {\em state space}. We define the ``flip action'', $\flip:V\times Q\to Q$, of a vertex on a switch position, as follows: $\flip(v,q)(u)=q(u)$ for all $u\in V\setminus\{v\}$ and $\flip(v,q)(v)=1-q(v)$, i.e., this action flips the function value of $q$ at $v$ only. We can then define a transition function $\delta: \Gamma \to \Gamma$ on a switch graph as $\delta((v,q))=(s^{q(v)}(v),\flip(v,q))$.

We define the {\em run} of a switch graph $G$ with initial state $\gamma_0:=(v_0,q_0)$ to be the unique infinite sequence over $\Gamma$,  $\Run[\infty]{G}{\gamma_0}:=(\gamma_i)_{i=0}^{\infty}$ satisfying $\gamma_{i+1}:=\delta(\gamma_i)$ for $i\geq 0$. For a vertex $v\in V$, we say a run {\em terminates at $v$} if $\exists t\in\nat$ such that $\forall i\geq t$  $\exists q_i\in Q$ with $\gamma_i=(v,q_i)$.  We call $T\in\natinf$ the {\em termination time} defined by $T:=\inf\{t\mid\forall i\geq t,\ v_i=v_t\}$, where $\inf\emptyset=\infty$. We denote by $\Run{G}{\gamma_0}:=(\gamma_i)_{i=0}^T$ the subsequence of $\Run[\infty]{G}{\gamma_0}$ up to the termination time $T$. We say a run {\em hits} a vertex $v\in V$ if $\exists t\in\nat$ and $\exists q_t\in Q$ such that $\gamma_t=(v,q_t)$. 

We note that in order to terminate at a vertex, $v\in V$, we must have that $v=s^0(v)=s^1(v)$. We  define the set 
of ``Dead Ends'' in the instance as $\DEnds:=\{v\in V \mid s^0(v)=s^1(v)=v\}$. From this definition, it is obvious that we either terminate at some unique vertex $v\in \DEnds$, or we never terminate. We may now define the Arrival Decision problem:

\problemStatement{Arrival}{\label{prob:sim:arrival}
Instance={A Switch Graph $G:=(V,s^0,s^1)$ and vertices $o,d\in V$.},
Problem={Decide whether or not the run of switch graph $G$ with initial state $(o,q^0)$ terminates at vertex $d$.}
}

Given a switch graph $G$ with directed edges, $E$, we define the relations $\arrowzero,\arrowone\subseteq V\times V$ as follows  $u\arrowzero v$ (resp. $\arrowone$) for $u,v\in V$ if and only if there is a directed path $w_0,\dots,w_k\in V$ with $(w_i,w_{i+1})\in E$ for $i\in[k-1]$, with $w_0=u$ and $w_k=v$ for $k\geq 0$ (resp. $k\geq 1$) from $u$ to $v$ in $(V,E)$. We write $u\not\arrowzero v$ (resp. $\not\arrowone$) whenever we do not have $u\arrowzero v$ (resp. $u\arrowone v$). 

We note that we can view the sequence of vertices visited on a run as a directed path in $(V,E)$, thus if the run with initial state $(v,q)$ hits $w$ then we can conclude $v\arrowzero w$ and, contrapositively, if $v\not\arrowzero w$ then for all $(v,q)\in Q$ the run starting at $(v,q)$ does not hit $w$.

We let $\mathbb{I}\{a=b\}$ be the indicator function of $a=b$, 
which is equal to $1$ if $a=b$ and is equal to $0$ otherwise. We now define a switching flow, rephrasing Definition 2 of Dohrau et al. \cite{DGKMW16}:

\begin{definition}[{\cite[Definition~2]{DGKMW16}}]\label{def:sim:edge-switching-flow}
Let $G:=(V,s^0,s^1)$ be a switch graph, and let $o,d\in V$ be vertices. We define a {\em switching flow} on $G$ from $o$ to $d$ as a vector $\bm{x}:=(x_e\mid e\in E)$ where $x_e\in\nat$ such that the following family of conditions hold for each $v\in V$:
\begin{alignat*}{3}
    &\mathrm{Flow\ Conservation:} \ \ && \left(\sum_{e=(u,v)\in E} x_e\right)-\left(\sum_{e=(v,w)\in E}x_e\right) = \mathbb{I}\{v=d\}-\mathbb{I}\{v=o\}, \ && \ \ \forall v\in V, \\
    &\mathrm{Parity\ Condition:} \ \  && x_{(v,s^1(v))}\leq x_{(v,s^0(v))}\leq x_{(v,s^1(v))}+1, \ && \ \ \forall v\in V.  
    \quad \quad \qed{}
\end{alignat*}
\end{definition}

We note that given $G$, $o$ and a switching flow $\bm{x}$ from $o$ to some, unknown, vertex $d\in V$, we can compute exactly which $d$ by verifying the equalities. We refer to $d$ as the {\em current-vertex} of the switching flow. If $o\in V$ is an initial vertex and $t\in\nat$ a time, we let $\Run{G}{(o,q^0)}:=((v_i,q_i))_{i=0}^\infty$ be the run, and define the {\em Run Profile} to time $t$ to be the vector $\RunPro{o,t}:=(\abs{\{i\in[t]\mid (v_{i-1},v_i)=e\}}\mid e\in E)$. It follows that for any $o\in V$ and $t\in\nat$ that $\RunPro{o,t}$ is a switching flow from $o$ to some vertex $d\in V$ \cite[Observation~1]{DGKMW16}. We say a switching flow $\bm{x}$ is {\em run-like} if there exists some $t\in\nat$ such that $\bm{x}=\RunPro{o,t}$.

It follows directly from the results of Dohrau et al.\cite{DGKMW16} and Gartner et al.\cite{GHH+18} that:

\begin{restatable}[\cite{DGKMW16,GHH+18}]{proposition}{simswitchflowwitness}\label{thm:sim:switch-flow-witness}
There exists a polynomial function $\mathrm{p}:\nat\to\nat$ such that for all Switch Graphs $G:=(V,s^0,s^1)$ and all vertices $o,d\in V$ with $o\neq d$ and $d\in\DEnds$ the following are equivalent:
\begin{itemize}
    \item The run on $G$ from initial state $(o,q^0)$ terminates at $d$.
    \item There exists a run-like switching flow $\bm{x}$ on $G$ from $o$ to $d$ satisfying $\forall e\in E$, that $\log_2(x_e) \leq \mathrm{p}(\abs{G})$.
\end{itemize}
Furthermore, for the same polynomial $\mathrm{p}$, the following are equivalent:
\begin{itemize}
    \item The run on $G$ from initial state $(o,q^0)$ does not terminate.
    \item There exists a vertex $d^\prime\in V\setminus\DEnds$, a run-like switching flow $\bm{x}$ on $G$ from $o$ to $d^\prime$, and an edge $e^\prime=(u,d^\prime)\in E$ which satisfies for all $e\in E\setminus\{e^\prime\}$ that $\log_2(x_e) \leq \mathrm{p}(\abs{G})$ and that $x_{e^\prime} = 2^{\mathrm{p}(\abs{G})}+1$.
\end{itemize}
\end{restatable}

It follows from these results that {\tt Arrival} is in $\NP\cap\coNP$, as we may non-deterministically guess a vector, $\bm{x}$, whose coordinate entries are bounded by $2^{\mathrm{p}(\abs{G})}+1$, and then verify whether or not $\bm{x}$ is a run-like switching flow. Using \cite[Lemma~11]{GHH+18} we may verify the run-like condition in polynomial time, on which we will elaborate subsequently. If we find a run-like switching flow to some dead end $d^\prime\in\DEnds$ we may conclude $G$ terminates at $d^\prime$ and by the first part of \Cref{thm:sim:switch-flow-witness} we can find such a flow within these bounds. This may be either a flow to the given dead-end $d$ in our input, or to some other dead-end, certifying non-termination at $d$. The last case
of \Cref{thm:sim:switch-flow-witness} says that when $G$ does not terminate anywhere, we may also find a flow certifying this within our bounds, namely with some coordinate value of the guessed vector $\bm{x}$ being exactly$2^{\mathrm{p}(\abs{G})}+1$ . In fact, it was shown by \cite{GHH+18} that this argument also shows containment 
of {\tt Arrival} in $\UP\cap\coUP$, by showing there is a unique witness
$\bm{x}$ satisfying just one of these conditions.

Let $G:=(V,s^0,s^1)$ be a Switch Graph and let $\bm{x}$ be a switching flow on $G$ between some vertices $o,d\in V$. We define the {\em last-used-edge} graph $\LueG{\bm{x}}:=(V,\LueE{\bm{x}})$ with the following set of edges:
\begin{align*}
    \LueE{\bm{x}}:=&\{(v,s^0(v))\mid  v\in V \text{ and } x_{(v,s^0(v))}\neq x_{(v,s^1(v))}\}\cup
                  \{(v,s^1(v))\mid  v\in V \text{ and } x_{(v,s^0(v))}= x_{(v,s^1(v))}>0\}
\end{align*}
This graph contains at most one of the edges $(v,s^0(v))$ or $(v,s^1(v))$.   If $x_{(v,s^0(v))}+ x_{(v,s^1(v))} > 0$, then
assuming there exists some run on which we visit vertex $v$ a total of $x_{(v,s^0(v))}+ x_{(v,s^1(v))}$ times, $\LueE{\bm{x}}$ contains the edge out of $v$ that our switching order would use the last time $v$ was visited on such a run. If on the other hand $x_{(v,s^0(v))}+ x_{(v,s^1(v))} = 0$,
then $\LueE{\bm{x}}$  contains neither edge.
%any run where we have never visited $v$ (i.e. $x_{(v,s^0(v))}+ %x_{(v,s^1(v))} =0$) then $\LueE{\bm{x}}$ contains neither edge, %since neither was ever used.

\begin{restatable}[{\cite{GHH+18}}]{proposition}{lastusededgecon}\label{sim:run-like-ver}
Let $G:=(V,s^0,s^1)$ be a Switch Graph and let $\bm{x}$ be a switching flow on $G$ from $o\in V$ to $d\in V$, then there exists a unique $t\in\nat$ such that $\bm{x}=\RunPro{o,t}$, if and only if one of the following two 
(mutually exclusive) conditions hold:
\begin{itemize}
    \item The graph $\LueG{\bm{x}}$ is acyclic,
    \item The graph $\LueG{\bm{x}}$ contains exactly one cycle and $d$ is on this cycle, 
\end{itemize}
Furthermore, given $G$ and any such $\bm{x}$ whether or not one of these conditions hold can be checked in polynomial time in the size of $G$ and $\bm{x}$.
\end{restatable}

\begin{restatable}[{\cite[Lemma~16]{GHH+18}}]{proposition}{lastusededgewelldef}\label{sim:last-used-edge-well-def}
Let $G:=(V,s^0,s^1)$ be a Switch Graph and let $t\in\nat$ with $\RunPro{o,t}$ the run profile up to time $t$, which is a switching flow on $G$ from $o\in V$ to some vertex $d\in V$. Then at least one of the following two conditions hold:
\begin{itemize}
    \item There is a unique edge $(u,d)\in\LueE{\RunPro{o,t}}$ incoming to $d$ in the graph $\LueG{\RunPro{o,t}}$.
    \item The graph $\LueG{\RunPro{o,t}}$ contains exactly one cycle,  and that cycle contains exactly one edge of the form $(u,d)\in \LueE{\RunPro{o,t}}$ on the cycle. 
\end{itemize}
Moreover, the edge $(u,d)$ was the edge traversed at time $t$ in the run (i.e., if $\Run[\infty]{G}{(o,q^0)}=((v_i,q_i))_{i=0}^{\infty}$ then $v_{t-1}=u$ and $v_t=d$).
Furthermore, this uniquely determined edge can be computed given $G$ and $\RunPro{o,t}$ in time polynomial in the size of $G$ and $\RunPro{o,t}$.
\end{restatable}

Using these results, we are able to efficiently (in \P{}-time) compute a function $\LUE$ which takes a switching flow of the form $\RunPro{o,t}$ and returns the ``last-used-edge'', namely the unique edge $(u,d)\in E$ guaranteed by \Cref{sim:last-used-edge-well-def}, where $(u,d)$ is the edge which was traversed at time $t$.

%% file: local-rec-arrival.tex
We consider a recursive generalisation of Arrival in the spirit of Recursive State Machines, etc. (\cite{AEY01,EY09,EY15}). A Recursive Arrival instance is defined as follows:

\begin{definition}\label{def:rec:rec-arrival-graph}
A {\em Recursive Arrival} graph is given by a tuple, $(G^1,\ldots,G^k)$, where each {\em component} $G^i:=(N_i\cup B_i, Y_i, \En_i, \Ex_i, \delta_i)$ consists of the following pieces:
\begin{itemize}
    \item A set $N_i$ of {\em nodes} and a (disjoint) set $B_i$ of {\em boxes}.
    \item A labelling $Y_i:B_i\to\{1,\ldots,k\}$ that assigns every box an index of one of the components $G^1,\ldots,G^k$.
    \item A set of {\em entry nodes} $\En_i\subseteq N_i$ and a set of {\em exit nodes} $\Ex_i\subseteq N_i$.
\item To each box $b \in B_i$, for all $i \in [k]$, 
we associate
a set of {\em call ports}, 
$\Call_{b,i} = \{(b, o) \mid 
o \in \En_{Y_i(b)} \}$ corresponding to the entries of
the corresponding component, and a set of {\em return ports}, $\Return_{b,i} = \{(b, d) \mid d \in
Ex_{Y_i(b)} \}$ corresponding to the exits of the corresponding component. We define the sets $\Call_i=\cup_{b\in B_i}\Call_{b,i}$ and $\Return_i=\cup_{b\in B_i}\Return_{b,i}$. We will use the term {\em ports} of $G^i$ to refer to 
the set $\Port_i = \Call_i \cup \Return_i$, of all call ports and return ports associated with all boxes $b \in B_i$ that occur
within the component $G^i$.
    
    \item A {\em transition relation}, $\delta_i$, where transitions are of the form $(u,\sigma,v)$ where:
    \begin{enumerate}
        \item The source $u$ is either a node in $N_i\setminus \Ex_i$ or a return port  
        $(b,x)$ in $\Return_i$.  We define $\Sor_i = N_i\setminus \Ex_i  \cup \Return_i$ to be the set of all {\em  source vertices}.
        \item The label $\sigma$ is either $0$ or $1$.
        \item The destination $v$ is either a node in $N_i\setminus \En_i$ or a call port $(b,e)$ where $b$ is a box in $B_i$ and $e$ is an entry node in $\En_j$ for $j=Y_i(b)$; we call this the set $\Dest_i$ of {\em destination vertices}.
    \end{enumerate}
    and we require that the relation $\delta_i$ has the following properties:
    \begin{enumerate}
        \item For every vertex $u\in\Sor_i$ and each $\sigma\in\{0,1\}$ there is a unique vertex $v\in \Dest_i$ with $(u,\sigma,v)\in\delta_i$. Thus, for each $i \in [k]$ and $\sigma \in \{0,1\}$, we can define total functions $s^\sigma_i:\Sor_i\to \Dest_i$ by the property that $(u,\sigma,s^\sigma_i(u))\in\delta_i$, for all $u \in \Sor_i$.\qed
    \end{enumerate}
\end{itemize}
\end{definition}

We will use the term {\em vertices} of $G^i$, which
we denote by $V_i$ to refer to the union $V_i = N_i \cup \Port_i$ of its set of nodes and its set of ports. For $\sigma \in \{0,1\}$, we let $E_i^\sigma = \{ (u,v) \mid (u,\sigma,v) \in \delta_i\}$ be the set of underlying edges of $\delta_i$ with label $\sigma$, and we define $E_i:=E_i^0\cup E_i^1$. We will often alternatively view components as being equivalently specified by the pair of functions $(s^0_i,s^1_i)$, which define the transition function $\delta_i:=\{(u,\sigma,s^\sigma_i(u))\mid u\in\Sor_i,\sigma\in\{0,1\}\}$.

We can view a box as a 
``call'' to other components, and, as such, it is natural to ask which components ``call'' other components. Given an instance of Recursive Arrival, $(G^1,\ldots,G^k)$, we define its {\em Call Graph} to be the following directed graph,  $C= ([k], E_C)$. Our vertices are component indices and for all $(i,j) \in [k] \times [k]$ let
$(i,j) \in E_C$ if and only if there exists some $b\in B_i$ with $j=Y_i(b)$ (i.e., a component $G^i$ can make a call to component $G^j$).  We allow self-loop edges in this directed graph, which correspond to a component making a call to itself.

We are also able to lift some definitions from non-recursive Arrival to analogous definitions about Recursive Arrival instances. Firstly, we define the sets $\DEnds_i:=\{v\in \Sor_i\mid s^0_i(v)=s^1_i(v)=v\}\cup\Ex_i$, of dead-ends of each component. This contains both vertices $v\in\Sor_i$ where both outgoing transitions are to itself and all the exits of the component.

In a given component, $G^i$, we define a {\em switch position} on $G^i$ as a function $q:\Sor_i\to\{0,1\}$. We let $Q_i$ be the set of all switch position functions on $G^i$. We let $q_i^0\in Q_i$ be the function $q_i^0(v)=0$ for all $v\in\Sor_i$ and call this the {\em initial switch position}. We define the action $\flip_i:\Sor_i\times Q_i\to Q_i$ analogously to non-recursive Arrival, which flips the bit corresponding to a given vertex in a given switch position.

A {\em state} of a Recursive Arrival graph $(G^1,\ldots,G^k)$ is given by a tuple $\gamma:=((b_1,q_1)\ldots(b_r,q_r),(v,q))$ where the {\em call stack} $\beta:=(b_1,q_1)\ldots(b_r,q_r)$ is a string of pairs $(b_i,q_i)$ with each $b_i\in\cup_k B_k$ a box, $q_i$ is a switch position on some component $G^{\cstack_i}$ (i.e. $q_i\in Q_{\cstack_i}$), and the {\em current position} is the pair $(v,q)$ where $v \in V_{\cstack_{r+1}}$ is a vertex in some component $G^{\cstack_{r+1}}$
and $q \in Q_{\cstack_{r+1}}$ is a switch position on $G^{\cstack_{r+1}}$. We call the sequence $(\cstack_1, \ldots, \cstack_r, \cstack_{r+1})$ the {\em component call-stack} of the state. We say that a state is {\em well-formed} if:
\begin{itemize}
    \item For all $i\in[r]$ we have $b_i\in B_{\cstack_i}$.
    \item The sequence satisfies $Y_{\cstack_i}(b_i)=\cstack_{i+1}$ for $i\in[r]$.
\end{itemize}
We let $\Gamma$ be the set of all well-formed states and $\Stacks:=\{\beta : \exists(v,q),\ (\beta,(v,q))\in\Gamma\}$ be the set of well-formed stacks $\beta$ appearing in some state of $\Gamma$.

We define the transition function $\delta:\Gamma\to\Gamma$ on a well-formed state $\gamma:=((b_1,q_1)\ldots(b_r,q_r),(v,q))$ as:
\begin{enumerate}
    \item If $v\in\Sor_j$ is a source vertex then we let $v^\prime:=s^{q(v)}(v)$ and then we define\\ $\delta(\gamma):=((b_1,q_1),\ldots,(b_r,q_r),(v^\prime,\flip_j(v,q)))$;
    \item If $v=(b,e)\in\Call_j$ then $e\in\En_{j^\prime}$ for $j^\prime=Y_j(b)$. We let $q^0_{j^\prime}$ be the initial switch position on $G^{j^\prime}$ and define $\delta(\gamma):=((b_1,q_1)\ldots(b_r,q_r)(b,q),(e,q^0_{j^\prime}))$;
    \item If $v\in\Ex_j$ and $r\geq 1$ then we define $\delta(\gamma):=((b_1,q_1)\ldots(b_{r-1},q_{r-1}),((b_r,v),q_r))$;
    \item If $v\in\Ex_j$ and $r=0$ then $\delta(\gamma):=\gamma$;
\end{enumerate}

The function $\delta:\Gamma\to\Gamma$ defines a deterministic transition system on well-formed states. We call the {\em run} of a Recursive Arrival graph from an initial component index $j\in[k]$, an initial switch position $q_0\in Q_j$ and a start entrance $o\in\En_j$ the (infinite) sequence $\Run[\infty]{G}{(o,q_0)}:=(\gamma_i)_{i=0}^\infty$ given by $\gamma_0:=(\epsilon, (o,q_0))$ and $\gamma_{i+1}:=\delta(\gamma_i)$. We say a run {\em terminates} at an exit $d\in\Ex_j$ if there $\exists t\in\nat$ such that $\forall i\geq t$ there $\exists q_i\in Q_j$ such that $\gamma_i=(\epsilon,(d,q_i))$. We call $T\in\natinf$ the termination time defined by $T:=\inf\{t\mid\forall i\geq t,\ v_i\in\Ex_j\}$, where $\inf(\emptyset)=\infty$. We denote by $\Run{G}{(o,q)}:=(\gamma_i)_{i=0}^T$ the subsequence up to termination. We say a run {\em hits} a vertex $v\in V$ if there $\exists t\in\nat$, $\exists q_t\in Q$ and $\exists\beta\in\Stacks$ with $\gamma_t=(\beta,(v,q_t))$.

Our decision problem can then be stated as:

\problemStatement{Recursive Arrival}{\label{prob:loc:gen-local-term-refined}
Instance={A Recursive Arrival graph $(G^1,\ldots,G^k)$, with $\abs{\En_j}=1$ for all $j\in[k]$, and a target exit $d\in\Ex_1$},
Problem={Does the run from initial state $(\epsilon,(o_1,q_1^0))$ terminate at exit $d$? (Where $o_1\in\En_1$ is the unique entry of $G^1$ and $q_1^0\in Q_1$ is the initial switch position.)}
}

This decision problem covers in full generality any termination
decision problem on Recursive Arrival instances, as we may accomplish a change of initial state by renumbering components and relabelling transitions. Also, restricting to models with $\abs{\En_i}=1$ is without loss of generality, 
because we can efficiently convert the model into an ``equivalent''
one where each component has a single entry,
by making copies of components (and boxes) with multiple entries, 
each copy associated with a single entry (single, call port, respectively).  This is 
analogous to the same fact for Recursive Markov Chains,
which was noted by Etessami and Yannakakis in \cite[p.~16]{EY09}. Thus, we may assume that in
the {\tt Recursive Arrival} problem
all components of the instance have a unique entry, i.e., for $i\in[k]$ that $\En_i=\{o_i\}$, and, unless stated otherwise, the run on $G$ refers to the run starting in the state $(\epsilon, (o_1,q_1^0))$, writing $\Run{G}{}:=\Run{G}{(o_1,q^0_1)}$. 

While, in such an instance, we may make an exponential number of calls to other functions, it turns out we are able to give a polynomial bound on the maximum recursion depth before we can conclude an instance must loop infinitely. 

\begin{restatable}{lemma}{recdepthbound}\label{cor:loc:rec-depth-bound}
Let $G:=(G^1,\ldots,G^k)$ be an instance of Recursive Arrival and assume the run on $G$ hits some state $(\beta,(v,q))$, with $\abs{\beta}\geq k$. Then
the run on $G$ does not terminate.    
\end{restatable}

\section{\P{}-Hardness of Recursive Arrival} \label{sec:p-hard}

Manuell \cite{Man21} has shown the {\tt Arrival} problem to be \PL{}-hard, which trivially provides the same hardness result for {\tt Recursive Arrival}.    This is currently the strongest
hardness result known for the {\tt Arrival} problem.
By contrast, we now show that the {\tt Recursive Arrival} problem is in fact \P{}-hard. 
\begin{restatable}{theorem}{twoexphard}\label{thm:loc:2ex-phard}
The {\tt 2-exit Recursive Arrival} problem is \P-hard.
\end{restatable}
\begin{proof}[Proof (Sketch)]
We show this by reduction from the P-complete Monotone Circuit Value Problem (see e.g., \cite{GHR95}). We construct one component corresponding to each gate of an input boolean circuit. Each component will have two exits, which we refer to as ``top'', $\top$, and ``bottom'', $\bot$, (located accordingly in our figures) and we will view these exits as encoding the outputs, ``true'' and ``false'' respectively. 

Firstly, we show in \Cref{fig:loc:2ex-phard-inputs} two components for a constant true and constant false gate of the circuit.
\begin{figure}
    \centering
    \begin{subfigure}[b]{0.45\textwidth}   
    \centering
    \resizebox{0.3\textwidth}{!}{\includegraphics{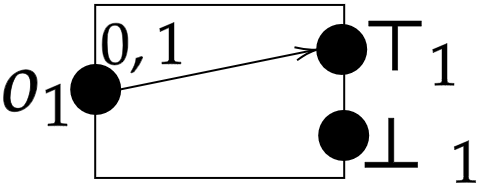}}
    \caption{$G^1$, the constant ``true'' component.}
    \label{fig:true-input}
    \end{subfigure}
    \hfill
    \begin{subfigure}[b]{0.45\textwidth}   
    \centering
    \resizebox{0.3\textwidth}{!}{\includegraphics{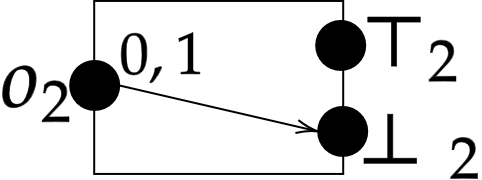}}
    \caption{$G^2$, the constant ``false'' component.}
    \label{fig:false-input}
    \end{subfigure}\caption{Initial components corresponding to constant gates ``true" and ``false". 
    %We observe $G^1$ always reaches exit $\top_{1}$ and $G^2$ %$\bot_{2}$. 
    %Edges labelled $0$ (resp. $1$) correspond to $s^0$ (resp. $s^1$) transitions.
    }
    \label{fig:loc:2ex-phard-inputs}
\end{figure}
Depicted in \Cref{fig:loc:2ex-phard-gates} are two cases corresponding to AND or OR gates. These perform a lazy evaluation of the AND or OR of components $G^j$ and $G^{k}$.
\begin{figure}
    \centering
    \begin{subfigure}[b]{0.47\textwidth}
        \centering
        \resizebox{0.8\textwidth}{!}{\includegraphics{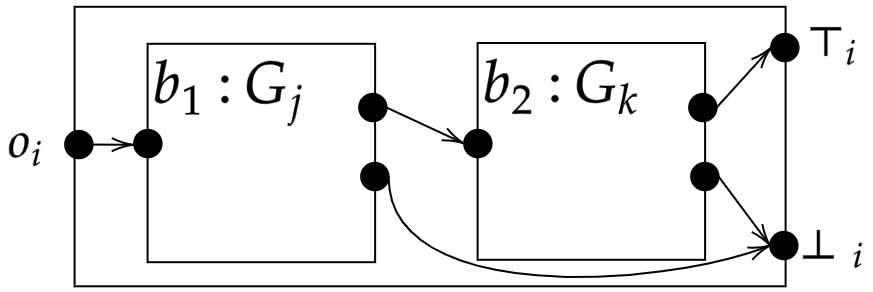}}
        \caption{Component for the AND of gates $g_j$ and $g_k$.}
        \label{fig:and-component}
    \end{subfigure}
    \begin{subfigure}[b]{0.47\textwidth}
        \centering
        \resizebox{0.8\textwidth}{!}{\includegraphics{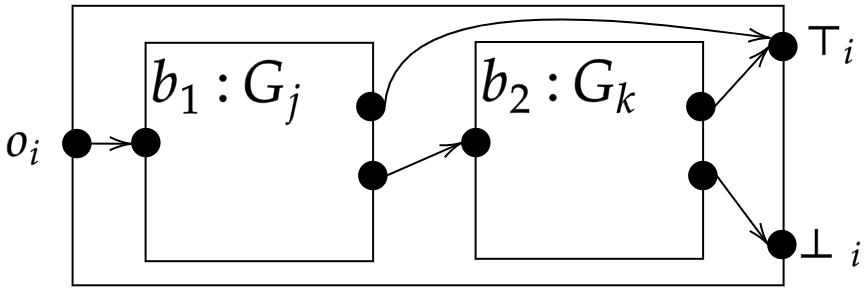}}
        \caption{Component for the OR of gates $g_j$ and $g_k$.}
        \label{fig:or-component}
    \end{subfigure}
    \hfill
    \caption{Component $G^i$, where $j,j^\prime\in[i-1]$ are the indices of the two inputs to the gate $g_{i}$. All edges correspond to both $s^0$ and $s^1$ transitions.}
    \label{fig:loc:2ex-phard-gates}
\end{figure}
This process produces a polynomially sized {\tt Recursive Arrival} instance for an input boolean circuit where each component $G_j$ can be shown inductively to reach exit $\top_j$ if and only if it's corresponding gate, $g_j$, outputs true.
\end{proof}

\section{Recursive Arrival is in $\NP{}\cap\coNP{}$ and \UEOPL{}}\label{sec:ueopl}
Recall the notion of Switching Flow for an Arrival instance.
For Recursive Arrival, we generalise the notion of a Switching Flow to a tuple of vectors $(\bm{x}^1,\ldots,\bm{x}^k)$, one for each
component of the Recursive Arrival instance. We define for each component $G^i$, $i\in [k]$, and each box $b\in B_i$ the set of {\em potential
edges} $F_{b,i}:=\Call_{b,i}\cross\Return_{b,i}$, representing the potential ways of crossing the box $b$, assuming that the
box is eventually returned from. We define the sets $F_i:=\cup_{b\in B_i}F_{b,i}$. We recall that the set of internal edges of
a component $G^i$ is given by $E_i:=\{(u,v)\mid u,v\in V_i,\ \exists\sigma\in\{0,1\}, (u,\sigma,v)\in\delta_i\}$. We say the {\em Flow Space} for component $G^i$ is the set of vectors $\mathcal{F}_i:={\nat}^{ |E_i \cup F_i|}:=\{(x^i_e \in \nat \mid e\in E_i\cup F_i)\}$, where we identify coordinates of these vectors with edges in $E_i\cup F_i$. We define the {\em Flow Space} of $G$ to be the set $\mathcal{F}:=\Pi_{i=1}^k\mathcal{F}_i$, a tuple of $k$ vectors, with the $i$'th vector in the  flow space of component $G^i$. We denote specifically by $\bm{0}^i\in\mathcal{F}_i$ the all zero vector, which has $\bm{0}^i_e=0$ for all $e\in E_i\cup F_i$, and $\bm{0}\in\mathcal{F}$ the all zero tuple, $\bm{0}:=(\bm{0}^1,\ldots,\bm{0}^k)$. We refer to elements of $\mathcal{F}$ (resp. $\mathcal{F}_i$) as flows on $G$ (resp. $G^i$).

Firstly, we define a switching flow on each component. For a Recursive Arrival instance $G:=(G^1,\ldots,G^k)$ and for $l\in[k]$, we call a vector $\bm{x}^l  \in\mathcal{F}_l$ to be a {\em component switching flow} if the following conditions hold.
Firstly, by definition, the all-zero vector $\bm{0}^l$
is always considered a component switching flow.
Furthermore, by definition, a non-zero vector $\bm{x}^l  \in\mathcal{F}_l\setminus \{ \bm{0}^l\}$ is called a component switching flow if there exists 
some 
{\em current-vertex}  $d^l_{\bm{x}^l}\in V_l \setminus \{ o_l \}$ (which, as we will see,
is always uniquely determined when it exists),
such that for $o_l$ the unique entry of $G^l$, $\bm{x}^l$ satisfies the following family of conditions: \\ \\
%for each $v\in V_l$ and $b\in B_l$:\\
$\begin{array}{cl}
    \mathrm{Flow\ Conservation} &\begin{cases}
    \left(\sum_{e=(u,v)\in E_l\cup F_l} x^l_e\right)&-\left(\sum_{e=(v,w)\in E_l\cup F_l}x^l_e\right) = 1,\qquad \mathrm{for }\ v=d^l_{\bm{x}^l},\\
   &+\left(\sum_{e=(v,w)\in E_l\cup F_l}x^l_e\right) = 1, \qquad \mathrm{for }\ v=o_l, \\
    \left(\sum_{e=(u,v)\in E_l\cup F_l} x^l_e\right)&-\left(\sum_{e=(v,w)\in E_l\cup F_l}x^l_e\right) = 0, \qquad \forall v\in V_l\setminus\{o_l,d^l_{\bm{x}^l}\},
    \end{cases}\\\\
    \mathrm{Switching\ Parity\ Condition} & \;\;\;\; x_{(v,s^1(v))}\leq x_{(v,s^0(v))}\leq x_{(v,s^1(v))}+1, \qquad\qquad\qquad\qquad \forall v\in \Sor_l,\\ \\
    \mathrm{Box\ Condition} & \;\;\;\; \exists f_b\in F_{b,l}$ such that $\forall f\in (F_{b,l} \setminus \{ f_b\}) \ \ x^l_f=0,  \qquad\qquad  \forall b \in B_l
    \end{array}
    $
\vspace*{0.1in}

Importantly, note that for any such component switching flow, $\bm{x}^l$, the current-vertex node $d^l_{x^l}$ is {\em uniquely} determined.  This follows
from the fact that the left-hand sides of the Flow
Conservation equalities for nodes $v \in V_l \setminus \{ o_l \}$ 
are identical and independent of the specific node $v$. Hence, if a vector $\bm{x}^l$ satisfies all
of those equalities, there can only be one
vertex $v \in V_l \setminus \{ o_l \}$ for which the corresponding
linear expression on the left-hand side,
evaluated over the coordinates of the vector $\bm{x}^l$, equals $1$.

In the case where $\bm{x}^l = \bm{0}^l$, i.e., the
all zero-vector, we define the current-vertex of the all-zero component switching flow to be $d^l_{\bm{0}^l}:=o_l$. We say a component switching flow $\bm{x}^l\in\mathcal{F}_l$ is {\em complete} if its current vertex $d^l_{\bm{x}^l}$ is an exit vertex in $\Ex_l$. These conditions follow the same structure as for non-recursive switching flows, with the additional ``Box Condition'' only allowing at most one potential edge across each box (i.e., an edge in $F_{b,l}$) to be used. 

Next, we extend our component switching flows by adding conditions that relate the flows on different components. Consider a tuple $\bm{X}:=(\bm{x}^1,\ldots,\bm{x}^k)\in\mathcal{F}$ of vectors, one for each component, such that each $\bm{x}^i\in\mathcal{F}_i$ is a component switching
flow for component $G^i$. We sometimes write $d^{i}_{\bm{X}}$ instead of $d^{i}_{\bm{x}^{i}}$.
Let $K_{\bm{X}} = \{ i \in [k] \mid \bm{x}^i \ \mbox{is complete}\}$ be the subset of indices  
corresponding to complete component switching flows. We then say the tuple $\bm{X}\in\mathcal{F}$ is a {\em recursive switching flow} if for 
every $l \in [k]$, $b\in B_l$ and $f\in F_{b,l}$, the following
holds:
\begin{itemize}
\item $\bm{x}^l \in \mathcal{F}_l$ is a component switching flow for component $G^l$, and  
\item if $x^l_f>0$ then $Y_l(b)\in K_{\bm{X}}$, and 
\item if $x^l_f > 0$,  then letting $d^{Y_l(b)}_{\bm{X}}\in\Ex_{Y_l(b)}$ be the current vertex of $\bm{x}^{Y_l(b)}$, we must have that $f=((b,o_{Y_l(b)}),(b,d^{Y_l(b)}_{\bm{X}}))$.
\end{itemize}
We define $\mathcal{R}\subset\mathcal{F}$ to be the set of all recursive switching flows. These conditions ensure ``consistency'' in the following way; if we use an edge $f\in F_{b,l}$ then we have a component switching flow on component $G^{Y_l(b)}$ which is complete and reaches the exit matching the edge $f$, and we are taking that same edge across all boxes with the same label.
We note our definition implies $\bm{0}\in\mathcal{R}$, thus there is always at least one recursive switching flow. These conditions can be verified in polynomial time.

\begin{figure}
    \centering
    \begin{subfigure}[b]{0.45\textwidth}
        \centering
        \resizebox{0.6\textwidth}{!}{\includegraphics{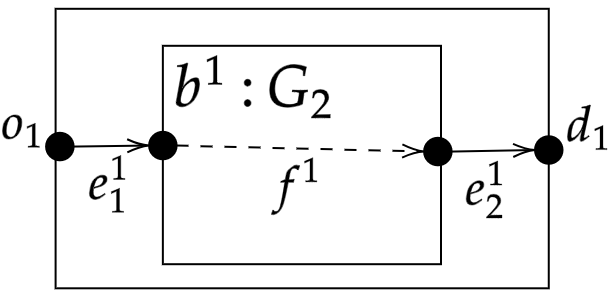}}
        \caption{Initial component $G^1$.}
        \label{fig:non-term:g1}
    \end{subfigure}
    \begin{subfigure}[b]{0.45\textwidth}
        \centering
        \resizebox{0.6\textwidth}{!}{\includegraphics{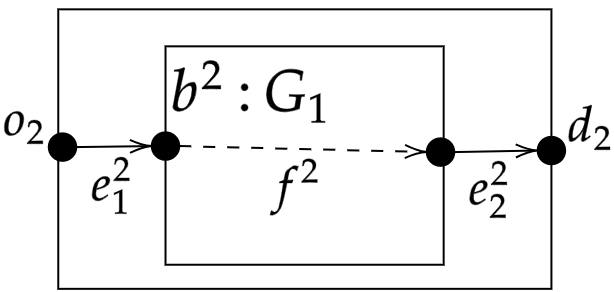}}
        \caption{Component $G^2$.}
        \label{fig:non-term:g2}
    \end{subfigure}
    \caption{A Recursive Arrival instance $G$ on which there exists a recursive switching flow $(\bm{x}^1,\bm{x}^2)$ on $G$ whose current vertex is in $G^1$ is $d_1$ however the run on $G$ does not terminate, or even hit the exit $d_1$. }
    \label{fig:non-term}
\end{figure}

We will view recursive switching flows as hypothetical partial ``runs'' on each component, where an edge $e\in E_l\cup F_l$ is used $x^l_e$ times along this ``run''. It may well be the case no such run actually exists. However, unlike
the case of non-recursive switching flows in Arrival, it is no longer the case that any recursive switching flow where the current vertex is $d_{\bm{X}}^1$
in component $G^1$, and where $d_{\bm{X}}^1 \in\Ex_1$ is an exit, 
necessarily certifies termination at $d_{\bm{X}}^1$.   It need not do so. For example, in the instance depicted in \Cref{fig:non-term} we may give the following flow on $G$: $\bm{x}^1=(1,1,1) \  ,\  \bm{x}^2=(1,1,1)$. The instance depicted obviously loops infinitely, alternating calls between components $G^1$ and $G^2$, but neither ever reaching an exit.  However,
the given $(\bm{x}^1,\bm{x}^2)$ corresponds to a recursive switching flow for this instance,  both of whose component switching flows have an exit as their current vertex. 

We need a way to determine whether the recursive switching flow
avoids such pathologies.   To do this, we need some additional definitions.
We describe a component switch flow $\bm{x}^l$ as {\em call-pending} if its current vertex $d^l_{\bm{x}^l}\in\Call_l$ is a call port, we let $J_{\bm{X}}\subseteq[k]$ be the set of all call-pending components and we let $r_{\bm{X}}:=\abs{J}$. From a recursive switching flow $\bm{X}:=(\bm{x}^1,\ldots,\bm{x}^k)$ we can compute the {\em pending-call graph} $\CallPenG{\bm{X}}:=([k],\CallPenE{\bm{X}})$ where we have edge $(i,j)\in\CallPenE{\bm{X}}$ if and only if $i\in J_{\bm{X}}$, $d_{\bm{X}}^i=(b,o)\in\Call_i$ is the current vertex of $\bm{x}^i$ and $j=Y_i(b)$. We can also compute the {\em completed-call graph}, 
$\ComCallG{\bm{X}}:=([k],\ComCallE{\bm{X}})$, where we have an edge $(i,j)\in\ComCallE{\bm{X}}$ if and only if $\exists b\in B_i,\ \exists f\in F_{i,b}$ with $x_f^i>0$ and $Y_i(b)=j$. The pending-call graph represents, from the perspective of an
imagined ``run'' corresponding to 
the recursive switching flow $\bm{X}$, which components $G^i$ are currently ``paused'' at a call port and waiting for component $G^j$ to reach an exit to determine the return port they should move to next. The completed-call graph represents the dependencies in the calls already made in 
such an imagined run, where an edge from component $G^i$ to component $G^j$ means that inside component $G^i$ the imagined run is making a
call to a box labelled by $G^j$ and 
``using'' the fact that component $G^j$, once called upon, reaches a specific exit.
In turn, in order to $G^j$ to reach its exit
the imagined run might be
``using'' the completion of other components to which there are outgoing edges from $G^j$ in
the completed-call graph. Thus, any cycle in the completed-call graph represents a series of circular (and hence not well-founded) assumptions about the imagined ``run'' corresponding to the recursive switching flow $\bm{X}$. For example, in the case of a 2-cycle between components $G^i$ and $G^j$, these are: ``If $G^i$ reaches exit $d_{\bm{X}}^i$ then $G^j$ reaches exit $d_{\bm{X}}^j$''; and ``If $G^j$ reaches exit $d_{\bm{X}}^j$ then $G^i$ reaches exit $d_{\bm{X}}^i$'' (c.f. \Cref{fig:non-term}).

Let $G$ be an instance of recursive arrival and let $\Run[\infty]{G}{o_1,q^0_1}:=(\beta_t,(v_t,q_t))_{t=0}^\infty$ be the run starting at $(o_1,q^0_1)$. We define the times $S_l:=\inf\{t\mid v_t=o_l\}$ and $T_l:=\inf\{t\mid v_t\in\Ex_l\}$ for each component index $l\in[k]$, with these values being $\infty$ if the set is empty. If $S_l<\infty$ we define the stack $\beta^l:=\beta_{S_l}$. We define the {\em component run} to be the (potentially finite) subsequence $t_1^l,t_2^l,\ldots$  of times which are precisely all times $t^l_j\in[S_l,\ldots,T_l]$ where $\beta_{t_l^j}=\beta^l$. We define the {\em Recursive Run Profile} of $G$ up to time $t$ as the sequence of vectors,  $\RunPros{G,t}:=(\RunPro{G^1,t}, \ldots, \RunPro{G^k,t})$, where for each $l \in [k]$,  $\RunPro{G^l,t}:=(\abs{ \{ j\in\nat\mid t^l_{j+1}\leq t \wedge (v_{t^{l}_{j}},v_{t^{l}_{j+1}})=e \} } \ \mid \ e\in E_l\cup F_l)$.    

In other words, $\RunPro{G^l,t}$ is a vector that provides counts of how many times each edge in component $G^l$ has been crossed, up to time $t$, during one ``visit'' to component $G^l$, with some particular call stack.   (The specific call stack doesn't matter.
This sequence does not depend on the specific calling context $\beta_l$ in which
$G^l$ was initially called.) We note that $\RunPro{G^l,0}=\bm{0}^l$.

Similarly to the non-recursive case, we can define the {\em last-used-edge} graph for each component $G^l$ as,  $\LueGr{l}{\bm{x}^l}:=(V_l,\LueEr{l}{\bm{x}^l})$ who's edge set is defined as:
\vspace*{-0.15in}
\begin{align*}
    \LueEr{l}{\bm{x}^l}:=&\{(v,s^0(v))\mid  v\in\Sor_l \text{ and } x^{l}_{(v,s^0(v))}\neq x^{l}_{(v,s^1(v))}\}\cup\\
        &\{(v,s^1(v))\mid  v\in\Sor_l \text{ and } x^{l}_{(v,s^0(v))}= x^{l}_{(v,s^1(v))}>0\} \ \cup \ 
        \{f\in F_l\mid x^l_f>0\}
\end{align*}

\vspace*{-0.1in}

We note that for the all-zero vector we have $\LueEr{l}{\bm{0}^l}=\emptyset$,  and if $\bm{x}^l\neq \bm{0}^l$ is non-zero then the current vertex $d_{\bm{x}^l}^l$ must have at least one incoming edge in $\LueEr{l}{\bm{x}^l}$, and thus the set $\LueEr{l}{\bm{x}^l}$ isn't empty.

Depending on how our run evolves, there are three possible cases:
\begin{itemize}
    \item For all $l \in [k]$, if $S_l<\infty$ then $T_l<\infty$.   This case corresponds to reaching some exit of $G^1$, i.e., terminating there.
    \item There exists some $l\in[k]$ with $S_l<\infty$ and yet with $T_l=\infty$,  however, where for all such $l\in[k]$ the subsequence $t^l_1,t^l_2,\ldots$ is of finite length.  This case corresponds to blowing up the call stack
    to arbitrarily large sizes, and as we shall describe, we can detect it by looking for a cycle in $\CallPenG{\bm{X}}$.
    \item There exists $l\in[k]$ with $S_l<\infty$ and $T_l=\infty$, where the subsequence $t^l_1,t^l_2,\ldots$ is of infinite length. This case corresponds to getting stuck inside component $G^l$, and infinitely often
    revisiting a vertex in a loop with the same call stack. As we shall see, we can detect this case by looking for a sufficiently large entry in some coordinate of $\bm{x}^l$.
\end{itemize}

Let $G$ be a Recursive Arrival instance and let $\bm{X}:=(\bm{x}^1,\ldots,\bm{x}^k)\in\mathcal{R}$ be a recursive switching flow on $G$, we say  $\bm{X}$ is {\em run-like} if it satisfies the following conditions:
\begin{itemize}
    \item For each component index $l\in[k]$ one of the following two conditions hold:
    \begin{itemize}
        \item The graph $\LueGr{l}{\bm{x}^l}$ is acyclic,
        \item The graph $\LueGr{l}{\bm{x}^l}$ contains exactly one cycle and $d^l_{\bm{x}^l}$ is on this cycle. 
    \end{itemize}
    \item If the set of call-pending component indexes $J_{\bm{X}}$ is non-empty, then $1\in J_{\bm{X}}$ and there is some total ordering $j_1,\ldots,j_{r_{\bm{X}}}$ of the set $J_{\bm{X}}$, with $j_1=1$, and a unique $j_{(r_{\bm{X}}+1)}\in[k]$ such that 
    the edges of the pending-call graph are given by $\CallPenE{\bm{X}} =\{(j_i,j_{i+1})\mid i\in [r_{\bm{X}}]\}$.   Note that we may have $j_{(r_{\bm{X}}+1)} = j_m$ for some $m \in \{1,\ldots,r_{\bm{X}}\}$, in which case $\CallPenE{\bm{X}}$ forms
    not a directed line graph but a ``lasso'' meaning a directed line ending in one directed cycle. When $J_{\bm{X}}=\emptyset$ we say that $r_{\bm{x}}:=0$ and that $j_1:=1$, thus the sequence is defined for all $\bm{X}$.
    \item For any $l\in[k]$ either: $l\in J_{\bm{X}}\cup K_{\bm{X}}$, or $\bm{x}^l=(0,\ldots,0)$, or $l=j_{(r_{\bm{X}}+1)}$.
    \item The completed-call graph $\ComCallG{\bm{X}}:=([k],\ComCallE{\bm{X}})$ is acyclic.
%    \item If the call-pending 
%graph $\CallPenG{\bm{X}}:=([k],\CallPenE{\bm{X}})$ %contains a cycle, that unique cycle must contain %the edge $(j_{r_{\bm{X}}},j_{(r_{\bm{X}}+1)})
%\in\CallPenE{\bm{X}}$.
    \item For any $l \in [k]$, if $\bm{x}^l \neq \bm{0}^l$, then
    in the graph $([k],\CallPenE{\bm{X}}
    \cup\ComCallE{\bm{X}})$ 
    we must have $1\arrowzero l$,
    i.e., there must be a path in this graph from component $1$ to all components $l$ for
    which $\bm{x}^l$ is non-zero.  
    %Equivalently,if $l\in K_{\bm{X}}$, then either $l=1$, or else $\exists i\in[k]$, $\exists b\in B_i$ with $Y_i(b)=l$,  $\exists(u,(b,o))\in E_i$, such that $x^i_{(u,(b,o))}>0$.
\end{itemize}
We denote by $\mathcal{X} \subset \mathcal{R}$ the set of all run-like recursive switching flows on $G$. We note for any $G$ that we always have $\bm{0}\in\mathcal{X}$.
%, all the graphs specified have no edges, $J_{\bm{X}}$ is empty, and all vectors are all zero, so this set is never empty
%. We will show that being run-like is equivalent to being equal to $\RunPros{G,t}$ for some $t\in\nat$ (\Cref{lem:run-like-characterization}).
We can show $\bm{X}\in\mathcal{F}$ is run-like if and only if $\exists t\in\nat$, $\bm{X}=\RunPros{G,t}$.

We now introduce ``unit vectors'' for this space, we write $\bm{u}^l_e\in\mathcal{F}_l$ for the vector where $u^l_e=1$ and for all other $e^\prime\in E_l\cup F_l$ with $e^\prime \neq e$ that $u^l_e=0$. We then write $\bm{U}_{i,e}\in\mathcal{F}$ for the sequence of $k$ vectors $(\bm{0}^1,\ldots,\bm{0}^{i-1},\bm{u}^i_e,\bm{0}^{i+1},\ldots,\bm{0}^k)$ where the $i$'th vector is $\bm{u}^i_e$ and for $i\neq j\in[k]$ that the $j$'th vector is the all-zero $\bm{0}^j$.  We may naturally define the notion of addition on $\mathcal{F}$ and we define the notion of subtraction $\bm{X}-\bm{U}_{i,e}$ in the natural way whenever $x^i_e>0$, i.e., the result of the subtraction remains in $\nat$ for every coordinate, subtraction is undefined where this isn't the case. We write $\mathcal{U}:=\{\bm{U}_{i,e}\mid i\in[k], e\in E_i\cup F_i\}$ for the set of all unit vectors.

Given a run-like recursive switching flow, $\bm{X}:=(\bm{x}^1,\ldots,\bm{x}^k)\in\mathcal{X}$, we say that $\bm{X}$ is {\em complete} if it is the case that $1\in K_{\bm{X}}$, i.e., the current vertex $d^1_{\bm{x}^1}$ of $\bm{x}^1$ is an exit of $G^1$. We say $\bm{X}$ is {\em lassoed} when $\CallPenE{\bm{X}}$ forms a ``lasso'', meaning a directed line ending in one directed cycle, as described earlier. We note that being complete and lassoed are mutually exclusive, because either $1 \in K_{\bm{X}}$ or $1 \in J_{\bm{X}}$, but not both.

\begin{restatable}{lemma}{uniqueincrsf}\label{lem:unique-inc-rsf}
    Let $G$ be an instance of Recursive Arrival, and let $\bm{X}\in\mathcal{X}$ be a run-like recursive switching flow on $G$. Then if $\bm{X}$ is neither complete nor lassoed, then there exists exactly one $\bm{U}_{i,e}\in\mathcal{U}$ such that $(\bm{X}+\bm{U}_{i,e})$ is a run-like recursive switching flow. Otherwise, if $\bm{X}$ is either complete or lassoed, then there exists no such $\bm{U}_{i,e}$.
\end{restatable}
\begin{proof}[Proof (Sketch)]
We shall show that for any $\bm{X}$ which is neither complete nor lassoed, we are able to give unique $i$ and $e$ as a function of $\bm{X}$. Viewing $\bm{X}$ as a ``hypothetical run'' to some time we use $J_{\bm{X}}$ as our ``call stack'' at this time and use that to determine the edge to increment. 
\begin{enumerate}
    \item If $d_{\bm{X}}^{j_{(r_{\bm{X}}+1)}}\in\Sor_{j_{(r_{\bm{X}}+1)}}$, then the ``current component'' is at a switching node and we take the edge given by our switching order. We note that this includes the case where $d_{\bm{X}}^{j_{(r_{\bm{X}}+1)}}=o_{j_{(r_{\bm{X}}+1)}}$, i.e. there is a call pending to a new component.
    \item If $j_{(r_{\bm{X}}+1)}\in K_{\bm{X}}$, then we can resolve the pending call in component $j_{r_{\bm{X}}}$ and increment the summary edge in $F_{j_{r_{\bm{X}}}}$ corresponding to exit $d_{\bm{X}}^{(j_{\bm{X}}+1)}$. 
\end{enumerate}
We can show that this is the unique choice in these cases through elimination, making use of the definitions of component, recursive, and run-like switching flows.
\end{proof}

We define the {\em completed call count} as the function $CC:\mathcal{F}\times [k]\to\nat$ which counts how many times a given component has been crossed
in a given flow, defined for $\bm{X}\in\mathcal{F}$ and $l\in[k]$ as follows:

\vspace*{-0.15in}

\begin{equation*}
    CC(\bm{X},l):=\sum_{i\in[k]} \quad \sum_{ \{b\in B_i \mid  Y_i(b)=l  \} }  \quad \sum_{f\in F_{b,i}} x^i_f
\end{equation*}

\vspace*{-0.05in}

\begin{restatable}{lemma}{uniquedecrsf}\label{lem:unique-dec-rsf}
    Let $G$ be an instance of Recursive Arrival, and let $\bm{X}\in\mathcal{X}$ be a run-like recursive switching flow on $G$. If $\bm{X}$ is non-zero then there exists a unique $\bm{U}_{i,e}\in\mathcal{U}$ such that $(\bm{X}-\bm{U}_{i,e}) \in \mathcal{X}$ is a run-like recursive switching flow. Otherwise, if $\bm{X}$ is all-zero, then no such $\bm{U}_{i,e}$ exists.
\end{restatable}
\begin{proof}[Proof (Sketch)]
We shall show for non-zero $\bm{X}$ the following choice is the unique value for $i$, and then $e$ can be determined using the last-used-edge graph in component $i$, as is the case for non-recursive switching flows. Viewing $\bm{X}$ as a ``hypothetical run'' to some time we use $J_{\bm{X}}$ as our ``call stack'' at this time and use that to determine the edge to decrement. 
\begin{itemize}
    \item If $\bm{x}^{j_{(r_{\bm{X}}+1)}}>\bm{0}^{j_{(r_{\bm{X}}+1)}}$ and $CC(\bm{X},j_{(r_{\bm{X}}+1)})=0$ then we decrement inside the ``current component'' as the pending-call in component $j_{r_{\bm{X}}}$ is the only call made.
    \item Otherwise, we take $i=j_{r_{\bm{X}}}$. Where, since we have either $CC(\bm{X},j_{(r_{\bm{X}}+1)})\geq 1$ or $\bm{x}^{j_{(r_{\bm{X}}+1)}}=\bm{0}^{j_{(r_{\bm{X}}+1)}}$ the current call from $j_{r_{\bm{X}}}$ to $j_{(r_{\bm{X}}+1)}$ is either made elsewhere and thus we cannot alter the component flow in $j_{(r_{\bm{X}}+1)}$ without affecting the edge traversed on these other calls or the flow in $j_{(r_{\bm{X}}+1)}$ is zero, in which case we step back from the final pending-call to it.
\end{itemize}
This can be shown to be the unique choice in each case through elimination.
\end{proof}

We define the function $Val:\mathcal{F}\to\nat$ as:
    $Val((\bm{x}^1,\ldots,\bm{x}^k)):=\sum_{i\in[k]}\sum_{e\in E_i\cup F_i} x^i_e$.
    This function sums all values across all vectors of the tuple. We note that for any flow $\bm{X}\in\mathcal{F}$ and any $i\in[k]$ and $e\in E_i\cup F_i$ that we have $Val(\bm{X}+\bm{U}_{i,e})=Val(\bm{X})+1$ and that when defined (i.e. $x^i_e>0$) that $Val(\bm{X}-\bm{U}_{i,e})=Val(\bm{X})-1$. 
    %Using \Cref{rec:run-pros-run-like,lem:unique-rsf-based-on-val} we have that $\mathcal{X}=\{\RunPros{G,t}\mid t\in\nat\}$ (\Cref{lem:run-like-characterization}).

Recall \Cref{thm:sim:switch-flow-witness} regarding non-recursive Arrival switching graphs, and in particular the fixed polynomial $\mathrm{p}$ which that proposition asserts the existence of. We say a recursive switching flow $\bm{X}:=(\bm{x}^1,\ldots,\bm{x}^k)\in\mathcal{X}$ is {\em finished} if it satisfies one of the following conditions:
\begin{enumerate}
    \item  $\bm{X}$ is complete, i.e, $1\in K_{\bm{X}}$, or, the current vertex $d_{\bm{X}}^1$ of $\bm{x}^1$ is an exit in $\Ex_1$.
    \item $\bm{X}$ is lassoed, i.e., $1\not\in K_{\bm{X}}$ and $j_{(r_{\bm{X}}+1)}\in J_{\bm{X}}$, or, the edges of $\CallPenE{\bm{X}}$ form a lasso.
    \item $\bm{X}$ is {\em just-overflowing}, which we define as follows:  $1\not\in K_{\bm{X}}$, and there exists some unique $l\in[k]$, 
    %with $d_{\bm{X}}^l$ the current-vertex of $\bm{x}^l$, 
    and unique $e=(u,d_{\bm{X}}^l)\in E_l\cup F_l$ with $x^l_e=2^{\mathrm{p}(\abs{V_l})}+1$, i.e., there is some unique component, $l$, and edge, $e$, incoming to its current vertex, $d^l_{\bm{X}}$, with a ``just-excessively large'' value in the flow $\bm{X}$. 
\end{enumerate}
We say the flow is {\em post-overflowing} if $1\not\in K_{\bm{X}}$, and there exists some $l\in[k]$, with $d_{\bm{X}}^l$ the current-vertex of $\bm{x}^l$, and some $e=(u,v)\in E_l\cup F_l$ satisfying at least one of: A) $x^l_e=2^{\mathrm{p}(\abs{V_l})}+1$ and $v\neq d_{\bm{X}}^l$; B) $x^l_e>2^{\mathrm{p}(\abs{V_l})}+1$.
We note that by repeatedly applying \Cref{lem:unique-dec-rsf} to a post-overflowing run-like recursive switching flow we must eventually find some finished just-overflowing run-like recursive switching flow.

We introduce the notation $\mathcal{F}^{N}\subseteq\mathcal{F}$ to be the restriction to tuples in which in every vector each coordinate is less than or equal to some $N\in\nat$. Thus $\mathcal{F}^N$ is finite, and any element $\bm{X}\in\mathcal{F}^N$ can be represented using at most $(\sum_{i=1}^k\abs{E_i\cup F_i})\cdot\log_2(N)$ bits. For all our subsequent results taking $N:=2^{\mathrm{p}(\max_l\abs{V_l})}+1$ will be sufficient, noting this means elements of $\mathcal{F}^N$ are represented using a polynomial number of bits in our input size. 

\begin{restatable}{theorem}{rainnpconp}\label{thm:ra-in-np-conp}
    The {\tt Recursive Arrival} problem is in $\NP{}\cap\coNP{}$ and $\UP{}\cap\coUP{}$.
\end{restatable}
\begin{proof}[Proof (Sketch)]
The proof follows from a series of lemmas given in the full version. These show: 
\begin{itemize}
    \item For any instance of {\tt Recursive Arrival}, $G$, there is a (unique) $\bm{X}\in\mathcal{F}^{N}$ which is a finished run-like recursive switching flow;
    \item Given any $\bm{X}\in\mathcal{F}^{N}$ we can verify whether or not $\bm{X}$ is a finished run-like recursive switching flow in \P{}-time;
    \item  Given any $\bm{X} \in \mathcal{F}^{N}$ which is a finished run-like recursive switching flow, we can determine whether or not $G$ terminates and if it does terminate at which exit in $\Ex_1$ it does so.
\end{itemize}
\end{proof}

\subsection{Containment in \UEOPL{}}
Given the previous results, we may consider a search version of {\tt Recursive Arrival} as follows:
\problemStatement{Search Recursive Arrival}{\label{prob:search-recursive-arrival}
Instance={A Recursive Arrival graph $(G^1,\ldots,G^k)$},
Problem={Compute the unique finished run-like recursive switching flow $(\bm{x}^1,\ldots,\bm{x}^k)\in\mathcal{F}$ on $G$}
}
In the appendix, we show that this problem is total and hence lies in $\TFNP{}$. We show containment in the total search complexity class $\UEOPL{}$ defined by Fearnley et al. \cite{FGMS19}, as problems polynomial time many-one search reducible to {\tt UniqueEOPL}, which is defined as follows:
\problemStatement{UniqueEOPL \cite{FGMS19}}{
Instance={Given boolean circuits $S,P:\{0,1\}^n\to\{0,1\}^n$ such that $P(0^n)=0^n\neq S(0^n)$ and a boolean circuit $V:\{0,1\}^n\to\{0,1,\ldots,2^{m}-1\}$ such that $V(0^n)=0$},
Problem={Compute one of the following:\begin{itemize}
    \item[(U1)] A point $x\in\{0,1\}^n$ such that $P(S(x))\neq x$.
    \item[(UV1)] A point $x\in\{0,1\}^n$ such that $x\neq S(x)$, $P(S(x))=x$, and $V(S(x))\leq V(x)$.
    \item[(UV2)] A point $x\in\{0,1\}^n$ such that $S(P(x))\neq x\neq 0^n$.
    \item[(UV3)] Two points $x,y\in\{0,1\}^n$, such that $x\neq y$, $x\neq S(x)$, $y\neq S(y)$, and either $V(x)=V(y)$ or $V(x)<V(y)<V(S(x))$.
\end{itemize}}
}
We may interpret an instance of {\tt UniqueEOPL} as describing an exponentially large directed graph in which our vertices are points $x\in\{0,1\}^n$ and each vertex has both in-degree and out-degree bounded by at most one. Edges are described by the circuits $S,P$, for a fixed vertex $x\in\{0,1\}^n$ there is an outgoing edge from $x$ to $S(x)$ if and only if $P(S(x))=x$ and an incoming edge to $x$ from $P(x)$ if and only if $S(P(x))=x$. We are given that $0^n$ is a point with an outgoing edge but no incoming edge or the ``start of the line''. We also have an ``odometer'' function, $V$, which has a minimal value at $0^n$. We assume our graph has the set-up of a single line $0^n,S(0^n),S(S(0^n)),\ldots$ along which the function $V$ strictly increases, with some ``isolated points'' where $x=S(x)=P(x)$. There are four types of solutions that can be returned, representing:
%the first represents the end of a line and the others are proofs of violations to one of our assumptions:
\begin{itemize}
    \item[(U1)] a point which is an ``end of the line'', with an incoming edge but no outgoing edge.
    \item[(UV1)] a violation of the assumption that valuation $V$ strictly increases along a line, since $V(x)\not<V(S(x))$.
    \item[(UV2)] a violation of the assumption there is a single line, since $x$ is the start of a line, but it is not $0^n$, thus it starts a distinct line.
    \item[(UV3)] a violation of one of the assumptions, however, in a more nuanced way. We can assume that $P(S(x))=x$ and $P(S(y))=y$, else they'd constitute a (UV1) example too, thus neither $x$ nor $y$ is isolated and both have an outgoing edge. If $x$ and $y$ were on the same line, then either $S(\ldots S(S(x)))=y$ or $S(\ldots S(y))=x$ by doing this iteration we'd eventually find some $z\in\{0,1\}^n$ where $V(z)\not< V(S(z))$, violating (UV1). However, if $x$ and $y$ are on different lines, then that would imply the existence of two distinct lines, violating (UV2). Thus, a (UV3) violation is a short proof of existence of a (UV1) or (UV2) violation elsewhere in the instance.   
\end{itemize}
For our reduction, our space will be made up of all possible flows $(\bm{x}^1,\ldots,\bm{x}^k)\in\mathcal{F}^{N}$ and our line will be made up of those arising from distinct $\RunPros{G,t}$'s, each step increasing in $t$ until we reach a finished flow, with all other vectors being isolated. A type (U1) solution will correspond to a finished run-like recursive switching flow, and we will show our instance has no (UV1-3) solutions, thus our computed solution to {\tt UniqueEOPL} will be a solution to {\tt Search Recursive Arrival}.

Given any flow $\bm{X}\in\mathcal{F}$ we can verify whether or not $\bm{X}$ is a run-like recursive switching flow (i.e. $\bm{X}\in\mathcal{X}\subset\mathcal{F}$). We will use this fact in our definitions of functions $Adv:\mathcal{F}\to\mathcal{F}$ and $Prev:\mathcal{F}\to\mathcal{F}$.

Our function $Adv$ on some value $\bm{X}:=(\bm{x}^1,\ldots,\bm{x}^k)\in\mathcal{F}$ is defined by the following sequence:
\begin{enumerate}
    \item If $\bm{X}\not\in\mathcal{X}$ then we take $Adv(\bm{X})=\bm{X}$. %Determine whether or not $\bm{X}\in\mathcal{X}$. 
    \item Else if $\bm{X}\in\mathcal{X}$ is either finished or post-overflowing then we take $Adv(\bm{X})=\bm{X}$.
    \item Otherwise, take $Adv(\bm{X})=\bm{X}+\bm{U}_{i,e}$, for the unique $\bm{U}_{i,e}\in\mathcal{U}$ such that $\bm{X}+\bm{U}_{i,e}\in\mathcal{X}$ (\Cref{lem:unique-inc-rsf}). 
\end{enumerate}
We note by this process that if $Adv(\bm{X})\neq\bm{X}$, then $Val(Adv(\bm{X}))=Val(\bm{X})+1$, since we have incremented exactly one edge in exactly one vector. Hence, this is consistent with our odometer.  We may also define the operation $Prev:\mathcal{F}\to\mathcal{F}$ analogously on some value $\bm{X}:=(\bm{x}^1,\ldots,\bm{x}^k)\in\mathcal{F}$. Taking $Prev(\bm{X})=\bm{X}$ whenever: $\bm{X}\not\in\mathcal{X}$; $\bm{X}=\bm{0}$, or; $\bm{X}$ is post-overflowing. Otherwise, taking $Prev(\bm{X})=\bm{X}-\bm{U}_{i,e}$, for the unique $\bm{U}_{i,e}\in\mathcal{U}$ such that $\bm{X}-\bm{U}_{i,e}\in\mathcal{X}$ (\Cref{lem:unique-dec-rsf}). Observe that, for any non-zero $\bm{X}\in\mathcal{X}$, that $Adv(Prev(\bm{X}))=\bm{X}$, and, for any $\bm{X}\in\mathcal{X}$, if we have $Prev(Adv(\bm{X}))\neq\bm{X}$, then $\bm{X}$ must be finished.

\vspace*{0.05in}

\begin{restatable}{theorem}{srecarrivalinueopl}
    The {\tt Search-Recursive Arrival} is in \UEOPL{}.
\end{restatable}

%\vspace*{-0.2in}

\begin{proof}[Proof (Sketch)]
We will give a polynomial-time search reduction from {\tt Search Recursive Arrival} to the {\tt UniqueEOPL} problem. We compute boolean circuits $S, P$ and $V$ which will be given by the restriction of the functions $Adv$, $Prev$, and $Val$ to the domain $\mathcal{F}^{N}$. This process involves computing membership of $\mathcal{X}$ and then computing the unique values $i$ and $e$ given by \Cref{lem:unique-inc-rsf,lem:unique-dec-rsf} for $Adv$ and $Prev$ respectively. We can then show that the only \UEOPL{} solution is of type (U1) and is a run-like recursive switching flow, which is a solution we are looking for.
\end{proof}

%By the 
%above results, we have placed {\tt Search Recursive Arrival} %inside \UEOPL{}, and using the known containment of Fearnley et %al. (\cite{FGMS19}) this places it in several well-known %subclasses of $\TFNP{}$ including $\CLS{}$ and, obviously thus, %both $\PLS{}$ and $\PPAD{}$. 

\vspace*{-0.2in}

\section{Conclusions}\label{sec:conclusions}

We have shown that {\tt Recursive Arrival} is contained in many of the same classes as the standard {\tt Arrival} problem. 
%Notably, the exception is the class \Tarski{}, of problems %reducible to computation of a fixed point of a monotone function %on a finite lattice. Containment of {\tt Arrival} in
%$\Tarski{}$ follows from the work of \cite{GHH21}, but containment %of {\tt Recursive Arrival} in \Tarski{} remains open. It turns out %that the example of \Cref{fig:non-term} provides one such hurdle %to \Tarski{} containment.
%if one were to try to 
%construct analogous monotone fixed point equations
%for the {\tt Recursive Arrival} problem, similar %to the
%equations
%for the {\tt Arrival} problem that arise from
%the work of \cite{GHH21}. 
While we have shown \P{}-hardness for {\tt Recursive Arrival}, whether or not {\tt Arrival} is \P{}-hard remains open.
%Our reduction from {\tt %Recursive Arrival} to the %search version of  
%{\tt Arrival} is a %polynomial-time Turing %reduction. Were it %possible to refine this %to, say, a logspace %reduction then this would %imply \P{}-hardness of %{\tt Arrival} as well.  

Let us note that the way we have chosen to generalise Arrival to the recursive
setting uses one of two possible natural choices for its semantics. 
Namely, it assumes a ``local'' semantics, meaning that the current switch position
for each component on the call stack 
is maintained as  part of the current state.
An alternative ``global'' semantics would instead consider the
switch position of each component as a ``global variable''. In such a model all switch positions would start in an initial position, and as the run progresses the switch positions would persist between, and be updated during, different calls to the same component. It is  possible to show (a result we have not included in this paper) that such a ``global" formulation immediately results in \PSPACE{}-hardness of reachability and termination problems.
%by reduction from the Quantified Boolean Formula (QBF) validity %problem. 

As mentioned in the introduction, a stochastic version of {\tt Arrival}, 
in which some nodes are switching nodes whereas other nodes are chance (probabilistic) nodes with probabilities on outgoing transitions, has already been studied in \cite{W22}, building on the work of \cite{FGMS19} which generalises {\tt Arrival} by allowing switching and player-controlled nodes. There is extensive prior work on RMCs and RMDPs,  with many 
known decidability/complexity results 
(see, e.g., \cite{EY09,EY15}).
It would be natural to ask
similar computational questions for
the generalisation of
RMCs and RMDPs
to a recursive Arrival model combining switching nodes with chance (probabilistic) nodes
and controlled/player nodes.

Finally, we note that Fearnley et al. also defined a \P{}-hard generalisation of {\tt Arrival} in \cite{FGMS19} which uses  ``succinct switching orders'' to succinctly encode an exponentially larger switch graph. We will refer to this 
problem as 
{\tt Succinct Arrival}. We don't know whether there are any
P-time reduction, in either direction, between {\tt Recursive Arrival} and {\tt Succinct Arrival}. It has been observed\footnote{Personal communication from Kousha Etessami and Mihalis Yannakakis.} that the results of \cite{GHH21} imply that both {\tt Arrival}
and {\tt Succinct Arrival} are P-time reducible to the {\tt Tarski}
problem defined in \cite{EPRY20}.  {\tt Succinct Arrival} is  also contained in \UEOPL{} by the same arguments as for {\tt Arrival}.
We do not currently know whether  {\tt Recursive Arrival} is 
P-time reducible to {\tt Tarski}.
%In the case of a reduction from {\tt Succinct Arrival} to {\tt %Recursive Arrival}, we note that each component of Recursive %Arrival is deterministic and on all calls reaches the same exit. %Thus it is unclear how one might use recursive components to %simulate a ``succinct node'' without the same simulation being %doable in {\tt Arrival}.  
%In the case of a reduction from {\tt Recursive Arrival} to {\tt %Succinct Arrival}, one somehow needs to use succinct nodes to %traverse the components recursively, detect if a cycle is %traversed and cache answers in each component, and it is unclear %how this can be achieved.

%As previously mentioned Fearnley et al. define 
%{\tt Succinct Arrival}, an alternate \P{}-hard generalisation of %{\tt Arrival}. We don't as yet know whether there is any P-time %reduction, in either direction, between {\tt Recursive Arrival} %and {\tt Succinct Arrival}. One can observe containment of {\tt %Succinct Arrival} in \Tarski{}, which is not known for {\tt %Recursive Arrival}. It is unclear how a reduction, in either %direction, since it would have to rely heavily on using ``succinct %nodes'' to mimic ``recursive calls'' (or v.v.) else imply a, as %yet unknown, P-time reduction from one of these generalisations to %{\tt Arrival}. 

\vspace*{0.1in}

\noindent {\bf Acknowledgements.}  Thanks to my PhD supervisor Kousha Etessami for his support.